\documentclass[10pt]{article}
\usepackage[letterpaper]{geometry}
\usepackage{hicss}
\usepackage{times}
\usepackage[none]{hyphenat}
\usepackage{url}
\usepackage{latexsym}
\usepackage{minted}
\usepackage{indentfirst}
\usepackage{graphicx}
\graphicspath{{images/}}
\usepackage[
    style=apa,
  ]{biblatex}
\input epsf
\usepackage{multirow}
\usepackage{balance} 
\usepackage{stfloats}
\usepackage{xcolor}

\definecolor{myyellow}{HTML}{FFF2CC}

\usepackage[english]{babel}
\usepackage[autostyle, english = american]{csquotes}
\MakeOuterQuote{"}

\usepackage{graphicx}
\usepackage{textcomp}
\usepackage{tcolorbox}
\usepackage{pifont}
\usepackage{xcolor}
\usepackage{soul}
\usepackage{booktabs}
\usepackage{multirow}
\usepackage{float}
\usepackage{url}
\usepackage{balance}
\usepackage{microtype}
\usepackage[numbered]{bookmark}
\usepackage{seqsplit}
\usepackage{newunicodechar,graphicx}
\usepackage[utf8]{inputenc}
\usepackage{pgf-pie}  
\usepackage{balance} 
\usepackage{algorithmic}
\usepackage{tabularx}
\usepackage[table,xcdraw]{xcolor}
\usepackage{array}
\usepackage{ragged2e}
\usepackage{adjustbox} 
\usepackage{makecell}
\usepackage{fontawesome5}

\newcommand{\RQA}{\textbf{RQ1}: To what extent do computing students understand OSS?
}
\newcommand{\RQB}{\textbf{RQ2}: What are the main barriers computing students face when onboarding to OSS projects?
}
\newcommand{\RQC}{\textbf{RQ3}: What are the enablers that computing students need to contribute to OSS development and maintenance?
}

\newcommand{\uniFull}{University of Hawai'i at Mānoa}
\newcommand{\deptFull}{Department of Information and Computer Sciences (ICS)}

\title{Enjoying the Product but Not Joining the Builders: Investigating Why Students Don’t Contribute to OSS

}

\author{
Italo Santos \\
University of Hawai'i at Mānoa \\
{\underline{isantos3@hawaii.edu}} \And
Briana Lee \\
University of Hawai'i at Mānoa \\
{\underline{brianall@hawaii.edu}} \And
Derek Garcia \\
University of Hawai'i at Mānoa \\
{\underline{dgarcia2@hawaii.edu}} \\
\AND
Yick Ching Wong \\
University of Hawai'i at Mānoa \\
{\underline{wongy@hawaii.edu}} \And
Anthony Peruma \\
University of Hawai'i at Mānoa \\
{\underline{peruma@hawaii.edu}}
}

\date{}

\begin{document}

\maketitle

\begin{abstract}
Open-source software (OSS) participation provides students with authentic exposure to collaborative, real-world software development, yet active contributions remain limited. This short paper presents an exploratory study of computing students’ perceptions, knowledge, and experiences with OSS at the \uniFull. We administered a survey and analyzed 122 responses using both quantitative and qualitative methods. Students reported moderate OSS familiarity and frequent OSS use in coursework and personal projects, yet only 12.3\% actively contributed and engagement with OSS communities was limited. Key barriers included skill gaps (72.89\% of non-contributors), time constraints (59.81\%), and codebase complexity (66.67\% of contributors), often alongside imposter syndrome (36.45\%). Students indicated that more time, mentorship, beginner-friendly issues, better onboarding/documentation, and curricular incentives (e.g., course credit) would most increase participation, while most reported their courses do not require or incentivize OSS contributions (58.20\%). Grounded in a single institutional context, this study provides a nuanced understanding of how curriculum structure, faculty encouragement, and competing academic demands shape OSS participation. These insights inform the design of targeted curricular and institutional interventions aimed at supporting student onboarding and participation in OSS communities.
\end{abstract}

\vspace{-1mm}
\subsubsection*{Keywords:}

open-source software, survey, software engineering education.

\vspace{-1mm}
\section{Introduction}
\label{intro}
\vspace{-3mm}

Software engineering projects are central to most computer science programs, and open-source software (OSS) plays an important role in preparing students for the software engineering workforce~(\cite{pinto2019training}). Unlike semester-long projects, OSS development enables students to engage in larger, community-driven projects that better reflect real-world software development, which involves longer histories and more complex technology stacks~(\cite{spinellis_why_2021, dekoenigsberg_how_2008}). While semester projects help students learn specific concepts, they often lack the practical experience needed in industry, contributing to skill gaps among graduates~(\cite{pinto2019training}). Participation in OSS further supports practical skill development and social learning, making it highly relevant in academia and industry~(\cite{salerno_barriers_2023}).

Although OSS offers significant educational and professional benefits, students, particularly those early in their careers, rarely participate as active contributors. This is striking given that students are well-positioned to engage in OSS due to their learning orientation and the wide range of contribution opportunities available, including non-coding tasks across diverse project domains and levels of complexity~(\cite{ellis2008support, von2003special, sarma2016training}). However, technical, social, and motivational barriers often hinder participation~(\cite{padala2020gender, trinkenreich2021women, cassee2026barriers}). Additionally, many students may not clearly perceive how OSS aligns with their academic and career goals, underscoring the need to better understand their perceptions and experiences within specific educational contexts.


Our aim is to characterize computer science students’ knowledge, prior experience, and perceptions of OSS, with a specific focus on understanding the persistent gap between OSS use and OSS contribution. Using an online survey, we measure students’ baseline awareness and conceptual understanding of OSS, identify the primary barriers that hinder onboarding into OSS projects, and determine which enablers would most effectively support sustained participation. We achieve our objective through the following research questions (RQs):




    \textbf{\RQA} Establishes a baseline of OSS knowledge and awareness among computer science students and determines whether additional exposure is needed.
    
    \textbf{\RQB} Identifies barriers that affect students’ participation and self-efficacy when engaging with OSS projects.
    
    \textbf{\RQC} Examines factors that enable positive onboarding experiences and encourage sustained student participation in OSS projects.



Our survey was administered to undergraduate and graduate students enrolled in computing-related degrees offered at the \uniFull. Through an analysis of 122 responses, we observe a consistent gap between attitudes and practices: students widely recognize the career and skill benefits of OSS. Respondents most frequently attributed non-participation to a lack of specific technical skills and insufficient time, while both contributors and non-contributors emphasized that additional time, mentorship, and structured onboarding resources (e.g., beginner-friendly issues, clearer documentation, and course credit) would make OSS contribution more feasible and sustainable.


The contribution of this study lies in offering a focused, context-specific examination of how computing students perceive, experience, and navigate OSS participation within a single institutional environment. Rather than proposing a new tool or intervention, this work establishes an empirical foundation for understanding the gap between OSS use and contribution, and identifies structurally embedded barriers and enablers that shape onboarding. Grounding the analysis in one academic context enables the generation of actionable insights to inform the design of targeted curricular and institutional interventions that better support student engagement with OSS communities.


\vspace{-2mm}
\section{Methodology}
\label{sec:method}
\vspace{-4mm}





\textbf{Study Design.} 
The survey included 29 questions (see Table~\ref{tab:survey}) that captured familiarity with OSS concepts and skills, experience contributing to OSS projects, challenges and obstacles to participating in OSS projects, and enablers and supports that help students feel more comfortable with OSS contributions. We utilized Qualtrics~\footnote{\url{https://www.qualtrics.com}} to design and distribute the survey, allowing each participant to provide one response.

\begin{table*}[!ht]
\scriptsize
\caption{OSS survey questions.}
\label{tab:survey}

\renewcommand{\arraystretch}{1.35}   
\setlength{\tabcolsep}{8pt}          

\begin{adjustbox}{width=\textwidth}
\begin{tabularx}{\textwidth}{>{\bfseries}c >{\raggedright\arraybackslash}X l >{\raggedright\arraybackslash}p{10em}}
\toprule
\textbf{\#} & \textbf{Survey Question} & \textbf{Type} & \textbf{Area} \\
\midrule
\rowcolor[HTML]{FFF2CC} 
1                               & Which of these best describes your gender identity?                                                                                & Single Choice    & \cellcolor[HTML]{FFF2CC}                               \\
\rowcolor[HTML]{FFF2CC} 
2                               & What best describes your ethnicity?                                                                                                & Multi-Choice     & \cellcolor[HTML]{FFF2CC}                               \\
\rowcolor[HTML]{FFF2CC} 
3                               & What is your age?                                                                                                                  & Single Choice    & \cellcolor[HTML]{FFF2CC}                               \\
\rowcolor[HTML]{FFF2CC} 
4                               & Which degree program are you currently enrolled in?                                                                                & Single Choice    & \cellcolor[HTML]{FFF2CC}                               \\
\rowcolor[HTML]{FFF2CC} 
5                               & What is your current academic standing?                                                                                            & Single Choice    & \cellcolor[HTML]{FFF2CC}                               \\
\rowcolor[HTML]{FFF2CC} 
6                               & How many years of programming experience do you have (including coursework and personal projects)?                                 & Single Choice    & \cellcolor[HTML]{FFF2CC}                               \\
\rowcolor[HTML]{FFF2CC} 
7                               & Have you had any software development internships, co-ops, or jobs?                                                                & Multi-Choice     & \multirow{-7}{*}{\cellcolor[HTML]{FFF2CC}Demographics} \\

\rowcolor[HTML]{C9DAF8} 
8                               & How familiar are you with the concept of open-source software (OSS)?                                                               & Single Choice    & \cellcolor[HTML]{C9DAF8}                               \\
\rowcolor[HTML]{C9DAF8} 
9                               & In your own words, how would you define “open source software”?                                                                    & Free Response    & \cellcolor[HTML]{C9DAF8}                               \\
\rowcolor[HTML]{C9DAF8} 
10                              & Where did you first learn about open-source software?                                                                              & Single Choice    & \cellcolor[HTML]{C9DAF8}                               \\
\rowcolor[HTML]{C9DAF8} 
11                              & Please provide the names of the open-source software tools, libraries, or applications that you use regularly:                     & Free Response    & \cellcolor[HTML]{C9DAF8}                               \\
\rowcolor[HTML]{C9DAF8} 
12                              & How often do you use OSS in different contexts? (coursework, personal projects, work, hackathons, clubs)                           & Single Choice    & \cellcolor[HTML]{C9DAF8}                               \\
\rowcolor[HTML]{C9DAF8} 
13                              & Rate agreement with statements about OSS importance: skill improvement, career value, quality, and community                       & Single Choice    & \cellcolor[HTML]{C9DAF8}                               \\
\rowcolor[HTML]{C9DAF8} 
14                              & Have you attended any OSS related events?                                                                                          & Single Choice    & \cellcolor[HTML]{C9DAF8}                               \\
\rowcolor[HTML]{C9DAF8} 
15                              & Have you ever contributed to an open source software project?                                                                      & Single Choice    & \cellcolor[HTML]{C9DAF8}                               \\
\rowcolor[HTML]{C9DAF8} 
16                              & How long have you been contributing to open source projects?                                                                       & Single Choice    & \cellcolor[HTML]{C9DAF8}                               \\
\rowcolor[HTML]{C9DAF8} 
17                              & In what ways have you contributed to OSS projects? (Check all that apply)                                                          & Multi-Choice     & \cellcolor[HTML]{C9DAF8}                               \\
\rowcolor[HTML]{C9DAF8} 
18                              & How would you rate your overall experience contributing to OSS projects?                                                         & Single Choice    & \multirow{-11}{*}{\cellcolor[HTML]{C9DAF8}RQ1}          \\ 


\rowcolor[HTML]{FFF2CC} 
19                              & Please describe any specific positive or negative experiences you had while contributing to OSS                                    & Free Response    & \cellcolor[HTML]{FFF2CC}                               \\
\rowcolor[HTML]{FFF2CC} 
20                              & What challenges have you encountered when contributing to OSS projects? (Check all that apply)                                     & Multi-Choice     & \cellcolor[HTML]{FFF2CC}                               \\
\rowcolor[HTML]{FFF2CC} 
21                              & What barriers prevent you from starting to contribute to OSS? (Check all that apply)                                               & Multi-Choice     & \cellcolor[HTML]{FFF2CC}                               \\
\rowcolor[HTML]{FFF2CC} 
22                              & Rank activities that compete with OSS contribution time. (coursework, exams, work, projects, clubs, social, self-care, job search) & Single Choice    & \multirow{-4}{*}{\cellcolor[HTML]{FFF2CC}RQ2}          \\

\rowcolor[HTML]{C9DAF8} 
23                              & What motivates you to contribute to OSS? (Check all that apply)                                                                    & Multi-Choice     & \cellcolor[HTML]{C9DAF8}                               \\
\rowcolor[HTML]{C9DAF8} 
24                              & What would encourage you to contribute more frequently to OSS? (Check all that apply)                                              & Multi-Choice     & \cellcolor[HTML]{C9DAF8}                               \\
\rowcolor[HTML]{C9DAF8} 
25                              & What would most motivate you to contribute to OSS? (Check all that apply)                                                          & Multi-Choice     & \cellcolor[HTML]{C9DAF8}                               \\
\rowcolor[HTML]{C9DAF8} 
26                              & Has a professor or instructor encouraged you to contribute to OSS?                                                                 & Single Choice    & \cellcolor[HTML]{C9DAF8}                               \\
\rowcolor[HTML]{C9DAF8} 
27                              & Have any of your courses required or incentivized OSS contributions?                                                               & Single Choice    & \cellcolor[HTML]{C9DAF8}                               \\
\rowcolor[HTML]{C9DAF8} 
28                              & Please briefly describe which course(s) and how OSS contributions were incorporated                                                & Free Response    & \cellcolor[HTML]{C9DAF8}                               \\
\rowcolor[HTML]{C9DAF8} 
29                              & What supports or resources could your university provide to help students contribute to OSS projects?                              & Multi-Choice     & \multirow{-7}{*}{\cellcolor[HTML]{C9DAF8}RQ3}         
\\

\bottomrule
\end{tabularx}
\end{adjustbox}
\end{table*}



\textbf{Pilot.} Before making our survey publicly available, we conducted a pilot to refine the instrument and ensure clarity and validity. The pilot involved five participants representative of our target population. Pilot participants completed the survey and provided feedback on question wording, structure, clarity, and overall flow. Based on their input, we rephrased ambiguous questions, reorganized question sequencing to improve coherence, and removed or refined questions that could lead to misinterpretation. Minor adjustments were also made to the response options to better capture students’ experiences and perceptions of OSS participation. Survey responses from the pilot were discarded and not included in our analysis.


\textbf{Sampling Strategy and Recruitment.} We used convenience sampling to recruit participants for our survey~(\cite{baltes_sampling_2022}). We connected with instructors in the \deptFull~ at \uniFull~ to distribute the survey among students in their class. 
The survey was open for three weeks, allowing ample time to gather sufficient feedback from the student body. This approach was suitable for our research, as our primary focus is on students' perceptions, barriers, and enablers to OSS participation. We included currently enrolled undergraduates and graduate students (Master’s and PhD level). We excluded students who are non-computing majors, alumni, and industry professionals. 




\textbf{Data Filtering.} We applied a filtering process to our responses to ensure data quality for analysis. This included removing submissions that were empty or incomplete, as well as responses from participants who indicated they were not enrolled in a computer science program. Following this filtering process, we were left with 122 valid responses for our analysis.


\textbf{Data Analysis.} We utilized a mixed-methods design that combined both quantitative and qualitative approaches~(\cite{wagner_challenges_2020}). Quantitative data were analyzed using descriptive and inferential statistics, while qualitative data from open-ended survey responses were examined through thematic analysis~(\cite{braun_using_2006}).  
Three authors independently coded the open-ended responses to identify recurring themes; any disagreements were resolved through discussion to reach a consensus. Individual responses could be assigned to multiple themes when appropriate. Responses that were ambiguous or not relevant to the research questions were excluded from the thematic analysis. Rather than calculating a quantitative measure of inter-rater reliability, we ensured quality through discussion. This approach to reaching agreement is known to effectively maintain coding quality in qualitative research~(\cite{Armstrong1997, Campbell2013}).


\textbf{Ethical Compliance.} Our study received Institutional Review Board approval. Informed consent was obtained from all participants, who were made aware of the study's purpose, voluntary participation, anonymity safeguards, and their withdrawal rights. No monetary compensation was provided to the participants.

\vspace{-2mm}
\section{Results}
\label{sec:results}
\vspace{-4mm}


In this section, we report the survey results by analyzing the responses to the 122 valid responses from students. We will first provide an overview of the participants' demographics, then address each RQ. In our RQ reporting, we begin by summarizing students' familiarity with OSS, their exposure to it, and the patterns of tool use and contributions (RQ1). Next, we discuss the barriers and competing demands on students' time that influence their onboarding experiences (RQ2). Finally, we explore the factors that motivate and support contributions from both contributors and non-contributors (RQ3). Where relevant, we complement our quantitative findings with illustrative quotes from open-ended responses.



\textbf{Participant Demographics.} Respondents predominantly identified as male (66.4\%, n=81) followed by female (30.3\%, n=37), with a small number identifying as non-binary/third gender (2.5\%, n=3) or preferring not to disclose (0.8\%, n=1). Regarding ethnicity, the most common groups were Asian (63.06\%, n=99) and White/Caucasian (19.11\%, n=30), followed by Native Hawaiian or Pacific Islander (7.64\%, n=12) and Hispanic or Latino (5.73\%, n=9), with smaller proportions reporting other identities. Most respondents were between 18--20 years old (40.16\%, n=49), with fewer participants aged 24 or older.


Participants primarily consisted of undergraduate computer science majors, including B.S. Computer Science (57.38\%, n=70), B.A. Computer Science (19.67\%, n=24), and B.S. Computer Engineering (4.92\%, n=6), alongside graduate students (M.S.: 13.93\%, n=17; Ph.D.: 4.10\%, n=5). 
Most respondents were third-year (36.89\%, n=45) or fourth-year students (19.67\%, n=24), followed by second-year students (15.6\%, n=19).

Respondents also reported varied programming backgrounds. Most indicated between 1--5 years of programming experience, with the largest group reporting 2--3 years (22.95\%, n=28), followed by 1--2 years and 3--4 years (21.31\%, n=26 each). A smaller proportion reported 4--5 years (17.21\%, n=21) or more than 5 years (14.75\%, n=18), and only a few reported less than one year (2.46\%, n=3). Finally, most respondents reported no prior software development internship, co-op, or job experience (64.75\%, n=79), while a smaller subset reported part-time roles or internships (13.93\%, n=17) or other forms of professional experience. Although our sample comes from a single institution, it includes students at different degree levels and academic standings, providing a suitable basis for examining perceptions and experiences of OSS participation. \\

\textbf{\RQA}

\begin{figure}[H]
    \centering
    \includegraphics[width=1.0\linewidth]{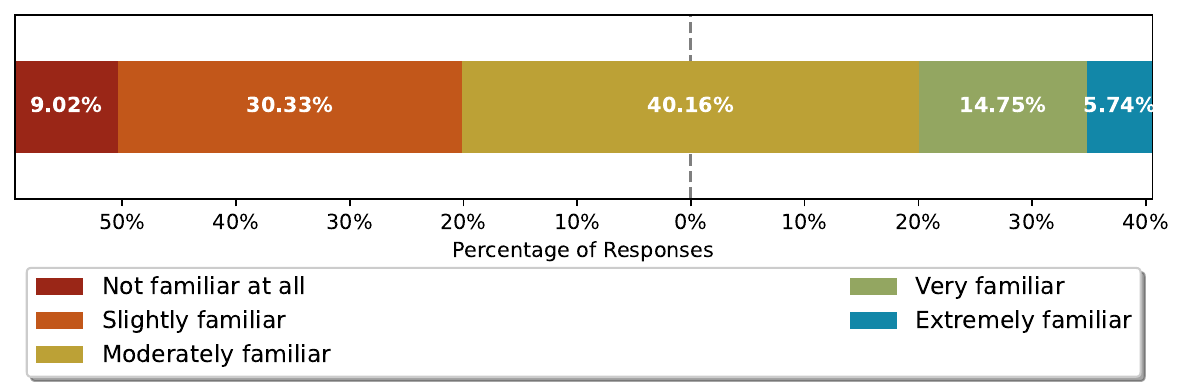}
    \caption{Self-reported familiarity with OSS among survey respondents (Question~\#8).}
    \label{fig:enter-label}
\end{figure}

To address this RQ, we examined students’ familiarity with OSS, sources of exposure, usage patterns, and contribution experiences. Overall, students demonstrated moderate familiarity with OSS concepts (Figure~\ref{fig:enter-label}). Specifically, 40.16\% (n=49) reported being ``moderately familiar,'' and 30.33\% (n=37) ``slightly familiar,'' while only 5.74\% (n=7) indicated being ``extremely familiar.'' When defining OSS in their own words, most emphasized public access and availability (62.64\%), with fewer mentioning modification rights (17.82\%) or community involvement (14.94\%), suggesting a primarily access-oriented understanding.

\begin{figure}[h]
    \centering
    \includegraphics[width=1\linewidth]{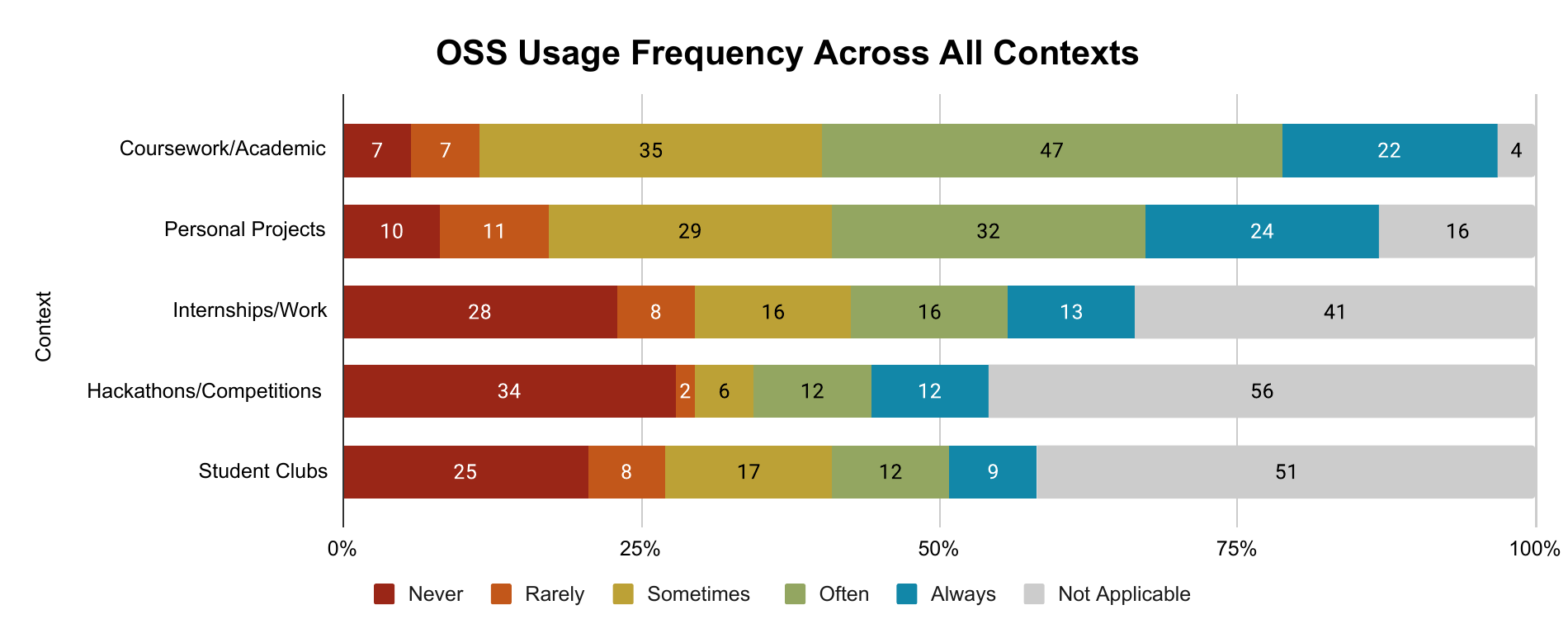}
    \caption{Distribution of responses to ``How often do you use OSS in the following contexts?''}
    \label{fig:context}
\end{figure}

University courses were the primary source of OSS exposure (32.79\%, n=40), followed by self-learning (13.93\%, n=17) and middle or high school classes (13.11\%, n=16). In practice, students most frequently reported using OSS in coursework and personal projects, while engagement in professional or extracurricular contexts remained limited (Figure~\ref{fig:context}). Attendance at OSS-related events was rare (88.89\%, n=112 reported no participation), indicating minimal interaction with the broader OSS community.

\begin{figure}[h]
    \centering
    \includegraphics[width=1\linewidth]{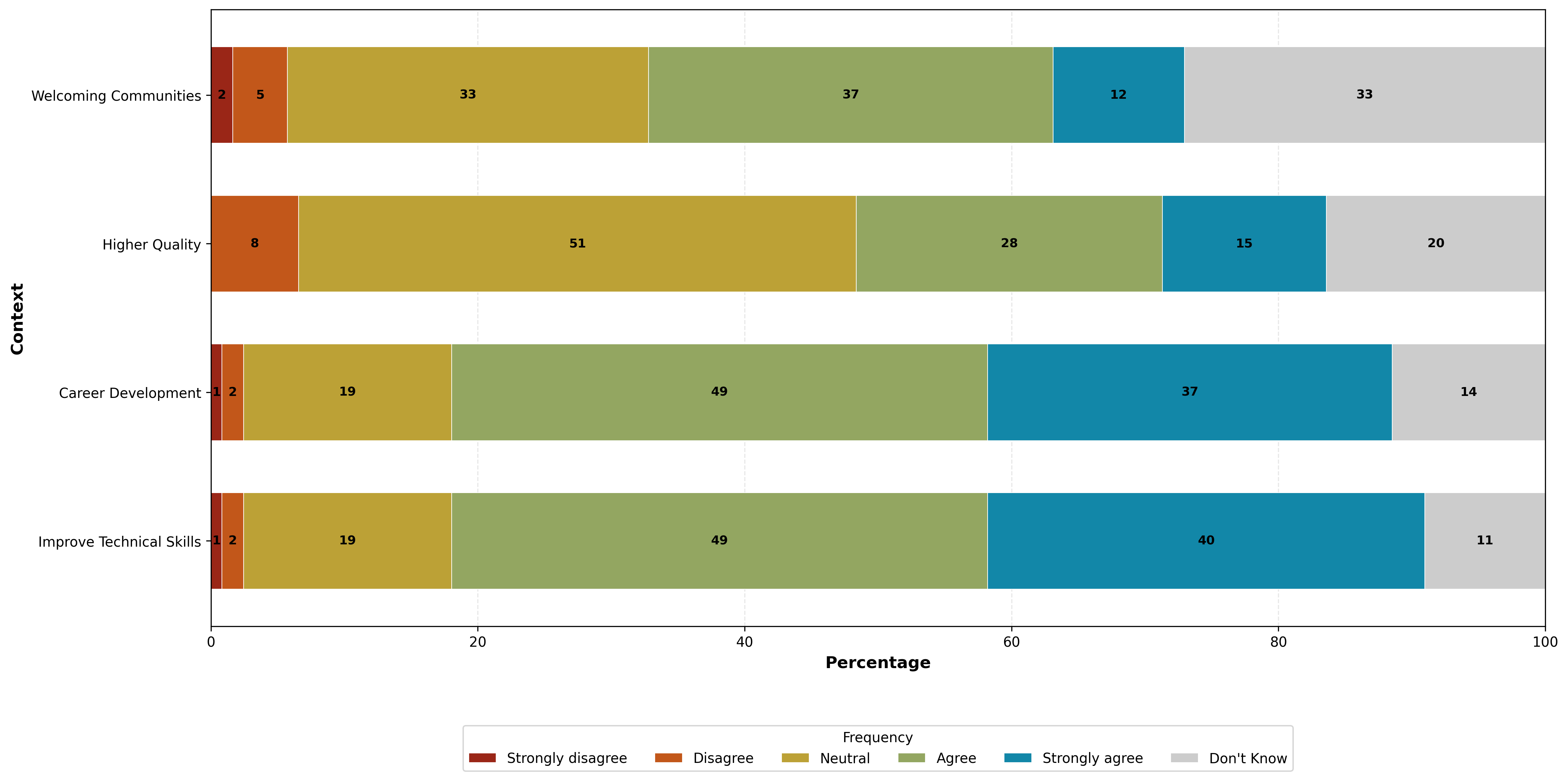}
    \caption{Distribution of OSS Experience}
    \label{fig:experience-areas}
\end{figure}

Students generally held positive attitudes toward OSS participation (Figure~\ref{fig:experience-areas}), with most agreeing that contributing would improve their technical skills and support career development. However, perceptions of OSS quality and community inclusiveness were mixed.

Despite regular OSS usage, only 12.3\% (n=15) of respondents had contributed to an OSS project. Descriptively, contribution rates were higher among students at more advanced academic levels, particularly among PhD respondents; however, the small subgroup sizes do not support broader inferential conclusions. Most contributors were relatively new and primarily engaged in code-centered activities such as submitting pull requests and opening issues. Overall, while students recognize the value of OSS, active contribution remains limited and largely disconnected from sustained community engagement.

\begin{tcolorbox}[colback=myyellow,colframe=black,top=0.5pt,bottom=0.5pt,left=1pt,right=1pt]
\noindent\faInfoCircle\hspace{0.3em}\textbf{\textit{Summary for RQ1.}}
Students demonstrated a moderate familiarity with OSS concepts, primarily understanding that OSS is publicly accessible (62.64\%). While most students regularly use OSS tools in coursework (56.56\% use often or always) and in personal projects (45.90\% use often or always), only 12.3\% of respondents actively contribute to OSS projects. Descriptively, contribution was more common among students at more advanced academic levels, although the small subgroup sizes limit interpretation. Students recognize the professional benefits and agree that it improves technical skills and career development. However, engagement with the broader OSS community is limited, and many have not attended OSS-related events.
\end{tcolorbox} 

\textbf{\RQB}

To address this RQ, we examined both the experiences of current contributors and the barriers preventing non-contributors from engaging in OSS. Among contributors (n=15), 11 provided responses about their overall contribution experience. Of these, six (54.54\%) described positive experiences and four (36.36\%) reported negative ones.


The qualitative responses further contextualize these challenges by showing that OSS participation can be difficult at the outset. Contributors described problems building projects, navigating undocumented abstractions, and learning unfamiliar tools and contribution practices. P003 noted that ``\textit{OSS is often very difficult to get into},'' while P069 described the experience as ``\textit{overwhelming when you first start}.'' P070 similarly reported using GitHub ``\textit{with practically no knowledge of Git}.'' At the same time, these difficulties could become learning opportunities: P044 explained that reading documentation and source code helped them ``\textit{learn very rapidly and improve my technical skills}.''

Contributors also reported mixed experiences with OSS communities and maintainers. Some described positive interactions, such as P004, who stated that ``\textit{the maintainer was very nice},'' and P031, who said that ``\textit{collaborating with others was mostly good}.'' Others experienced delays, limited engagement, or governance challenges; for example, P001 reported that ``\textit{it took 6 months for maybe 10 lines of code to be merged},'' while P002 cited a ``\textit{lack of community engagement and timeliness}.'' Despite these challenges, participants such as P020 emphasized that contributing was ``\textit{super valuable and I learned a lot}'' suggesting that OSS participation can simultaneously present technical and social barriers while providing meaningful opportunities for skill development and collaboration.

Among contributors (n=15), the most frequently reported challenges were understanding complex codebases (n=10, 66.67\%), lacking specific technical skills (n=8, 53.33\%), and difficulty identifying beginner-friendly issues (n=8, 53.33\%). These barriers often co-occurred, suggesting that onboarding challenges compound rather than occur in isolation.

Barriers were more pronounced among non-contributors (n=107). The most significant obstacle was lack of specific technical skills (n=78, 72.89\%), followed by insufficient time (n=64, 59.81\%), imposter syndrome (n=39, 36.45\%), and difficulty understanding complex codebases (n=37, 34.58\%). Compared to contributors, non-contributors reported stronger skill- and time-related constraints, indicating that these factors play a central role in preventing initial participation.


We used the Borda Count method~(\cite{saari_selecting_2023}) to determine the overall ranking, as per Table~\ref{tab:ranking}, the primary indicator of competing activities was ``coursework and assignments,'' ranked \#1, followed by ``studying for exams'' and ``paid work or internships.'' A further analysis of rankings by degree program revealed that undergraduates and master's students ranked ``coursework and assignments'' as \#1, ``studying for exams'' as \#2, and ``paid work or internships'' as \#3. Whereas PhD students ranked ``paid work or internships'' as \#1, ``coursework and assignments'' as \#2, and ``personal coding projects'' as \#3. For contributors and non-contributors, the rankings were identical for the top three: ``coursework and assignments,'' ``studying for exams,'' and ``paid work or internships.'' Starting at rank 4, contributors prioritized ``personal coding projects,'' ``social/family activities,'' ``rest and self-care,'' ``job search and career preparation,'' and ``extracurricular activities or clubs.'' Non-contributors starting at rank 4 prioritized ``social/family activities,'' ``rest and self-care,'' ``extracurricular activities or clubs,'' ``job searching and career preparation,'' and ``personal coding projects.''

\begin{table}[!h]
\caption{Most potential to contribute to OSS is ranked by the Borda count method~\cite{saari_selecting_2023}.}
\label{tab:ranking}
\resizebox{\columnwidth}{!}{%
\begin{tabular}{llll}
\hline
\textbf{Rank} & \textbf{Answer Option}               & \textbf{Borda Score} & \textbf{Prop} \\ \hline
1             & Coursework and assignments           & 913                  & 1.52          \\ \hline
2             & Studying for exams                   & 752                  & 1.25          \\ \hline
3             & Paid work or internships             & 596                  & 0.99          \\ \hline
4             & Social/Family activities             & 470                  & 0.78          \\ \hline
5             & Rest and self-care                   & 431                  & 0.72          \\ \hline
6             & Extracurricular activities or clubs  & 415                  & 0.69          \\ \hline
7             & Job searching and career preparation & 411                  & 0.69          \\ \hline
8             & Personal coding projects             & 404                  & 0.67          \\ \hline
\end{tabular}%
}
\end{table} 

\begin{tcolorbox}[colback=myyellow,colframe=black,top=0.5pt,bottom=0.5pt,left=1pt,right=1pt]
\noindent\faInfoCircle\hspace{0.3em}\textbf{\textit{Summary for RQ2.}}

Students faced multiple barriers when onboarding to OSS projects. Current contributors most often struggled with understanding complex codebases (66.67\%), lack of specific technical skills (53.33\%), and difficulty finding beginner-friendly issues (53.33\%). These barriers often cluster together, with 46.67\% of contributors experiencing both codebase complexity and difficulty finding suitable first issues. Non-contributors faced barriers: 72.89\% cited a lack of technical skills, and 59.81\% reported insufficient time. Imposter syndrome affected 36.45\% of non-contributors. Competing time demands primarily stem from coursework and assignments (ranked \#1 overall), studying for exams (\#2), and paid work or internships (\#3), though PhD students prioritize paid work over coursework. Contributors' experiences were mixed, with 54.54\% reporting positive experiences and 36.36\% describing negative experiences. 
\end{tcolorbox}

\textbf{\RQC}

To address this RQ, we examined motivations and factors that could increase student participation in OSS. Among contributors (n=15), the primary motivations were building a professional portfolio (n=9, 60.00\%) and improving programming skills (n=8, 53.33\%), often combined with meeting academic requirements. Contributors reported that more free time (n=11, 73.33\%), mentorship from experienced contributors (n=9, 60.00\%), and better onboarding documentation (n=8, 53.33\%) would encourage more frequent contributions.

Among non-contributors (n=107), the most cited factors that would increase participation were more free time (n=69, 64.49\%), mentorship from experienced contributors or faculty (n=65, 60.74\%), beginner-friendly issues (n=56, 52.34\%), and academic course credit (n=56, 52.34\%). Financial compensation was selected by n=38 (35.51\%).

Institutional support appeared limited. Nearly half of respondents (n=56, 45.90\%) reported no faculty encouragement to contribute to OSS, and 58.20\% (n=71) indicated that their courses did not require or incentivize OSS contributions. Descriptively, reported faculty encouragement was more frequent at higher academic levels, particularly among PhD respondents; however, the small graduate subgroup sizes limit broader interpretation of this pattern. Overall, the results suggest that time availability, mentorship, structured onboarding, and curricular integration are central factors influencing students’ engagement with OSS.

\begin{tcolorbox}[colback=myyellow,colframe=black,top=0.5pt,bottom=0.5pt,left=1pt,right=1pt]
\noindent\faInfoCircle\hspace{0.3em}\textbf{\textit{Summary for RQ3.}}
Several key factors enable students to contribute to OSS. For non-contributors the most critical enabler was \enquote{more free time} (64.49\%), followed by \enquote{mentorship from experienced contributors or faculty} (60.74\%), and \enquote{more beginner-friendly issues} and \enquote{academic courses credit} (both 52.33\%). Current contributors similarly prioritize \enquote{more free time} (73.33\%) and \enquote{mentorship from experienced contributors} (60.00\%), along with \enquote{better onboarding documentation} (53.33\%). The motivations for both groups focused on \enquote{building professional portfolios} for 60\% of contributors and 59.81\% of non-contributors, and \enquote{improving programming skills} motivating 53.33\% of contributors and 53.27\% of non-contributors. 
\end{tcolorbox}

\section{Discussion and Implications}
\label{sec:discussion}
\vspace{-4mm}

Our findings reinforce several barriers already documented in OSS newcomer research, including technical skill gaps, complex codebases, difficulty identifying suitable first issues, and limited onboarding support~(\cite{steinmacher2015systematic, santos2022hits, santos2025software}). Rather than suggesting that these barriers are unique to students in our sample, our results help explain how they interact with the educational environment in which students encounter OSS. Three related tensions are particularly evident in our findings: a gap between OSS use and contribution, a gap between motivation and the feasibility of participation, and a mismatch between students' available time and the way OSS contribution is positioned within the curriculum.

\textbf{OSS use does not necessarily translate into OSS contribution.}
Students in our sample were regularly exposed to OSS and frequently used OSS in coursework and personal projects, yet only 12.30\% (n=15) reported having contributed to an OSS project. Students also generally perceived OSS contribution as beneficial for developing technical skills and supporting their careers. These findings suggest an important distinction between being an OSS \textit{user} and becoming an OSS \textit{participant}. Familiarity with OSS tools and exposure through coursework can introduce students to OSS as a software-development resource without necessarily providing the situated knowledge required to navigate an unfamiliar repository, identify an appropriate task, interact with maintainers, and complete a contribution. This distinction is consistent with prior work showing that newcomer onboarding involves multiple technical and social barriers beyond initial awareness or interest~(\cite{padala2020gender, cassee2026barriers}). In our context, therefore, increased exposure to OSS alone appears insufficient to bridge the transition from consuming OSS to participating in its development.

\textbf{Students' motivation to contribute is constrained by the feasibility of doing so.} Although students viewed OSS contribution as valuable for developing technical skills and supporting their careers, participation was constrained by practical barriers. Non-contributors most often reported limited technical skills and time, while contributors emphasized codebase complexity, skill gaps, and difficulty finding beginner-friendly issues. Similar technical and development challenges have been reported among OSS developers more broadly~(\cite{peruma2026understanding}), but may be amplified for students by limited experience and academic demands. These barriers can compound one another, as skill uncertainty may increase the time needed to understand a codebase, identify a feasible task, and make a contribution. Thus, motivation alone may be insufficient without the time, skills, and support needed to participate.

\textbf{Students face a curriculum--time mismatch when attempting to contribute.}
Time emerged as one of the clearest links between individual barriers and the educational environment. Students ranked coursework and assignments as their primary competing demand, followed by studying for exams and paid work or internships. At the same time, 58.20\% of respondents reported that their courses did not require or incentivize OSS contributions, and nearly half reported receiving no faculty encouragement to contribute. These findings reveal a structural tension: students perceive OSS participation as educationally and professionally valuable, but when contribution is not incorporated into coursework, it may require additional time beyond the academic activities that already dominate their schedules. Consequently, asking students to contribute to OSS without integrating the activity into existing coursework may effectively position contribution as additional extracurricular work.

This curriculum--time mismatch provides a context-specific explanation for why ``more free time'' emerged as the most frequently selected enabler among both contributors and non-contributors. The issue may therefore not be resolved simply by encouraging students to devote additional personal time to OSS. Instead, the findings suggest that educational structures can influence whether students have realistic opportunities to participate. Curricular integration, mentorship, structured onboarding, and carefully selected entry-level tasks may reduce the additional cost of participation by embedding OSS activity within students' existing learning responsibilities. In this sense, our results complement prior newcomer research by illustrating how established technical and social barriers may be amplified or mitigated by the institutional context in which students encounter them.

Taken together, these three tensions help characterize the gap between students enjoying and benefiting from OSS products and becoming active participants in their development. Our results do not suggest that technical barriers, time constraints, or onboarding challenges are unique to this population. Instead, the contribution of this study is to show how these established barriers intersect within an educational setting: students may have exposure without a pathway to participation, motivation without sufficient opportunity or support, and interest in contributing while academic structures compete for the time required to do so. This perspective shifts the educational challenge from simply increasing awareness of OSS toward creating feasible and supported pathways from OSS use to OSS contribution.


\textbf{Implications for educators and OSS communities.}
Our findings suggest three priorities for supporting OSS participation. First, \textit{integrating OSS contributions into coursework} can reduce time constraints while emphasizing the contribution process rather than pull-request acceptance~(\cite{pinto2019training, holmes2018dimensions}). Second, \textit{structured mentorship and onboarding} can reduce technical and procedural barriers. Third, educators should \textit{curate beginner-appropriate projects and issues} aligned with students' skills~(\cite{fronchetti2019attracts}, Santos et al.~\citeyear{santos2023designing}). These interventions could also be personalized to students' skills and experience, building on prior work exploring conversational AI tailored to different problem-solving styles~(Santos et al.~\citeyear{santos2025great}).

\textbf{Implications for research and the value of a single-context lens.}
Where prior studies often examine multiple communities or propose onboarding solutions at the project level~(\cite{PS02canfora2012going, park2009beyond}, Santos et al.~\citeyear{santos2025software}), our study complements this literature by examining how established OSS barriers and enablers manifest within a specific educational environment. The institutional context is important for interpreting these findings. Our sample consisted predominantly of undergraduate computing students; most respondents reported limited prior professional software-development experience; university courses represented an important source of OSS exposure; and formal curricular incentives for OSS contribution were limited. These characteristics may partly explain why coursework emerged as the strongest competing demand and why students placed substantial value on mentorship, structured onboarding, and course credit.

Accordingly, we do not interpret the observed frequencies as population-level estimates of computing students more broadly. Instead, the findings support analytical generalization by illustrating how motivation, technical preparedness, available time, and institutional support can interact to shape OSS participation within an educational setting. Replication across institutions with different student populations, curricular structures, levels of OSS integration, and geographic contexts is needed to determine which patterns persist across settings. Within our context, students' stated motivation to contribute coexists with strong feasibility constraints, suggesting that interventions should address both students' motivation to participate and the opportunities and support available for doing so.

\section{Threats to Validity}
\label{sec:threats}
\vspace{-4mm}

This study has limitations that constrain the interpretation and transferability of the findings~(\cite{ralph2020empirical}). Data collection relied on non-probability sampling and self-reported responses from students at a single institution, introducing potential risks of sampling bias, recall bias, and social desirability effects. Participation was voluntary, and students with stronger opinions or greater interest in OSS may have been more likely to respond, potentially affecting representativeness.

The findings are therefore suitable for analytical generalization rather than population-level inference. Although the sample includes students across degree levels and academic standings, the results may not generalize to institutions with different student populations, curricular structures, professional-development opportunities, or levels of OSS integration. In particular, the predominance of undergraduate students, limited professional software-development experience among many respondents, and relatively limited curricular incentives for OSS contribution may have influenced the prominence of time constraints, mentorship needs, and requests for course credit observed in our results. Future replications across diverse institutions and geographic contexts are needed to determine which patterns remain consistent across educational settings.

To mitigate internal validity concerns, we conducted a pilot study, applied screening and filtering procedures, and used multiple data quality checks. Open-ended responses were independently coded by three authors, with disagreements resolved through discussion to reduce subjectivity. Nevertheless, interpretation of qualitative responses remains inherently subjective.

Construct validity may also be affected by variation in how students interpreted certain survey terms. Although we took care in survey design and piloting, differences in interpretation may affect measurement accuracy. We also avoid inferential comparisons across several academic-level and contribution-status subgroups because some groups, particularly the contributor and graduate subgroups, contain few respondents. Accordingly, we report subgroup differences descriptively rather than making inferential comparisons.

Finally, the study is descriptive and exploratory. While we report patterns related to OSS familiarity, barriers, and motivations, the design does not support causal claims. The findings should therefore be understood as reflecting students' reported experiences within this institutional context rather than as definitive evidence of broader causal relationships.

\section{Related Work}
\label{sec:relatedwork}
\vspace{-4mm}


\textbf{Onboarding in OSS.} Newcomers often struggle to understand complex codebases, identify suitable first issues, and navigate asynchronous collaboration environments~(\cite{steinmacher2015systematic, carter_skillscope_2025}). Issue trackers may lack beginner-friendly guidance, while mature projects often rely on tacit knowledge that is difficult for students to access. 


\textbf{Supporting Student Participation.} Prior work has proposed tools and interventions to ease OSS onboarding. Automated approaches have been used to characterize issues based on the technical domains associated with the APIs involved in their implementation~(Santos et al.~\citeyear{santos2023tag}), while others, such as SkillScope, predict the skills required for OSS issues~(\cite{carter_skillscope_2025}). Research also shows that motivation and structured educational support can encourage participation. For example, intrinsic motivation is associated with greater interest in OSS contribution~(Santos et al.~\citeyear{santos_exploring_2025}), while gamified environments such as OSSDoorway can improve student self-efficacy and engagement~(Santos et al.~\citeyear{santos_ossdoorway_2025}).


\textbf{Barriers and Inclusion.} Significant barriers nevertheless remain. Students frequently report difficulties with code comprehension, technical skills, and documentation quality~(\cite{salerno_barriers_2023}). Underrepresented groups may face additional socio-cultural barriers, including discrimination, toxic environments, imposter syndrome, and limited representation in leadership~(\cite{trinkenreich2021women}).

\textbf{Positioning of This Work.} While prior studies have identified common onboarding barriers, motivations, and support mechanisms, many focus on broad populations or specific interventions. In contrast, our study provides a contextualized investigation of students' familiarity, experiences, barriers, and motivations within a single institutional environment. This perspective allows us to examine how curriculum, faculty encouragement, and local context shape OSS participation and can inform targeted educational strategies.

\section{Conclusions and Future Work}
\label{sec:conclusion}
\vspace{-4mm}



OSS plays a critical role in preparing students for professional software engineering practice, yet meaningful participation remains limited. Students face persistent challenges when attempting to contribute, including identifying suitable projects, setting up development environments, and navigating complex or insufficient documentation. These challenges can limit students' opportunities to gain authentic
software development experience during their academic training, resulting in missed opportunities for experiential learning.

To better understand this gap, we conducted an exploratory survey within the ICS department at \uniFull, collecting responses from 122 students. Rather than seeking broad generalizations, our goal was to investigate a single institutional context in depth to uncover how curriculum structure, faculty encouragement, local culture, and time pressures shape students’ OSS perceptions and behaviors. This focused lens allows us to identify context-specific barriers and opportunities that may be obscured in large cross-institutional studies.

Our findings reveal a clear tension: students recognize the professional and skill-development value of OSS, yet most lack direct experience contributing to OSS and encounter structural and motivational barriers. These insights provide a foundation for future work, including the design and evaluation of targeted curricular and institutional interventions to better support students' onboarding into OSS communities. Future interventions will be grounded in a detailed understanding of this specific educational context, enabling the development of scalable yet context-aware strategies that strengthen the connection between classroom learning and authentic collaborative software development.

\vspace{-3mm}
\section*{Data Availability}
\vspace{-4mm}
The data and materials used in this study are publicly available at:
\url{https://github.com/dlg1206/uh-manoa-ics-oss-survey}.




\printbibliography

@article{ralph2020empirical,
  title={Empirical standards for software engineering research},
  author={Ralph, Paul and Ali, Nauman bin and Baltes, Sebastian and Bianculli, Domenico and Diaz, Jessica and Dittrich, Yvonne and Ernst, Neil and Felderer, Michael and Feldt, Robert and Filieri, Antonio and others},
  journal={arXiv preprint arXiv:2010.03525},
  year={2020}
}

@article{Armstrong1997,
  title = {The Place of Inter-Rater Reliability in Qualitative Research: An Empirical Study},
  volume = {31},
  ISSN = {1469-8684},
  url = {http://dx.doi.org/10.1177/0038038597031003015},
  DOI = {10.1177/0038038597031003015},
  number = {3},
  journal = {Sociology},
  publisher = {SAGE Publications},
  author = {Armstrong,  David and Gosling,  Ann and Weinman,  John and Marteau,  Theresa},
  year = {1997},
  month = aug,
  pages = {597–606}
}

@article{Campbell2013,
  title = {Coding In-depth Semistructured Interviews: Problems of Unitization and Intercoder Reliability and Agreement},
  volume = {42},
  ISSN = {1552-8294},
  url = {http://dx.doi.org/10.1177/0049124113500475},
  DOI = {10.1177/0049124113500475},
  number = {3},
  journal = {Sociological Methods\&; Research},
  publisher = {SAGE Publications},
  author = {Campbell,  John L. and Quincy,  Charles and Osserman,  Jordan and Pedersen,  Ove K.},
  year = {2013},
  month = aug,
  pages = {294–320}
}

@article{santos2023tag,
  title={Tag that issue: applying API-domain labels in issue tracking systems},
  author={Santos, Fabio and Vargovich, Joseph and Trinkenreich, Bianca and Santos, Italo and Penney, Jacob and Britto, Ricardo and Pimentel, Jo{\~a}o Felipe and Wiese, Igor and Steinmacher, Igor and Sarma, Anita and others},
  journal={Empirical Software Engineering},
  volume={28},
  number={5},
  pages={116},
  year={2023},
  publisher={Springer}
}

@inproceedings{peruma2026understanding,
  title={Understanding npm Developers’ Practices, Challenges, and Recommendations for Secure Package Development},
  author={Peruma, Anthony and Choy, Truman and Lee, Gerald and Santos, Italo De Oliveira},
  booktitle={Proceedings of the 19th International Conference on Cooperative and Human Aspects of Software Engineering},
  %pages={272--282},
  year={2026}
}

@inproceedings{santos2025great,
  title={Great power brings great responsibility: Personalizing conversational ai for diverse problem-solvers},
  author={Santos, Italo and Felizardo, Katia Romero and Steinmacher, Igor and Gerosa, Marco A},
  booktitle={2025 IEEE/ACM 18th International Conference on Cooperative and Human Aspects of Software Engineering (CHASE)},
  pages={93--95},
  year={2025},
  organization={IEEE}
}

@article{santos2025software,
  title={Software solutions for newcomers’ onboarding in software projects: A systematic literature review},
  author={Santos, Italo and Felizardo, Katia Romero and Steinmacher, Igor and Gerosa, Marco A},
  journal={Information and Software Technology},
  volume={177},
  pages={107568},
  year={2025},
  publisher={Elsevier}
}

@inproceedings{cassee2026barriers,
  title={What barriers do students experience when trying to contribute to Open Source Software projects?},
  author={Cassee, Nathan and Ravi, Sankarsh and Santos, Italo and Steinmacher, Igor and Serebrenik, Alexander},
  booktitle={Proceedings of the IEEE/ACM 48th International Conference on Software Engineering},
  pages={151--161},
  year={2026}
}

@inproceedings{dekoenigsberg_how_2008,
	title = {How {Successful} {Open} {Source} {Projects} {Work}, and {How} and {Why} to {Introduce} {Students} to the {Open} {Source} {World}},
	url = {https://ieeexplore.ieee.org/document/4556977/},
	doi = {10.1109/CSEET.2008.42},
	urldate = {2025-12-12},
	booktitle = {2008 21st {Conference} on {Software} {Engineering} {Education} and {Training}},
	author = {DeKoenigsberg, Gregory},
	month = apr,
	year = {2008},
	note = {ISSN: 1093-0175},
	pages = {274--276},
}

@article{spinellis_why_2021,
	title = {Why computing students should contribute to open source software projects},
	volume = {64},
	issn = {0001-0782, 1557-7317},
	url = {https://dl.acm.org/doi/10.1145/3437254},
	doi = {10.1145/3437254},
	language = {en},
	number = {7},
	urldate = {2025-12-12},
	journal = {Communications of the ACM},
	author = {Spinellis, Diomidis},
	month = jul,
	year = {2021},
	pages = {36--38},
}

@article{saari_selecting_2023,
	title = {Selecting a voting method: the case for the {Borda} count},
	volume = {34},
	issn = {1043-4062, 1572-9966},
	shorttitle = {Selecting a voting method},
	url = {https://link.springer.com/10.1007/s10602-022-09380-y},
	doi = {10.1007/s10602-022-09380-y},
	language = {en},
	number = {3},
	urldate = {2025-11-30},
	journal = {Constitutional Political Economy},
	author = {Saari, Donald G.},
	month = sep,
	year = {2023},
	pages = {357--366},
}

@article{braun_using_2006,
	title = {Using thematic analysis in psychology},
	volume = {3},
	issn = {1478-0887, 1478-0895},
	url = {https://www.tandfonline.com/doi/full/10.1191/1478088706qp063oa},
	doi = {10.1191/1478088706qp063oa},
	language = {en},
	number = {2},
	urldate = {2025-10-06},
	journal = {Qualitative Research in Psychology},
	author = {Braun, Virginia and Clarke, Victoria},
	month = jan,
	year = {2006},
	pages = {77--101},
}

@misc{wagner_challenges_2020,
	title = {Challenges in {Survey} {Research}},
	url = {http://arxiv.org/abs/1908.05899},
	doi = {10.48550/arXiv.1908.05899},
	urldate = {2025-09-16},
	publisher = {arXiv},
	author = {Wagner, Stefan and Fernández, Daniel Méndez and Felderer, Michael and Graziotin, Daniel and Kalinowski, Marcos},
	month = apr,
	year = {2020},
	note = {arXiv:1908.05899},
}

@article{baltes_sampling_2022,
	title = {Sampling in software engineering research: a critical review and guidelines},
	volume = {27},
	issn = {1573-7616},
	shorttitle = {Sampling in software engineering research},
	url = {https://doi.org/10.1007/s10664-021-10072-8},
	doi = {10.1007/s10664-021-10072-8},
	language = {en},
	number = {4},
	urldate = {2025-09-16},
	journal = {Empirical Software Engineering},
	author = {Baltes, Sebastian and Ralph, Paul},
	month = apr,
	year = {2022},
	pages = {94},
}

@misc{salerno_barriers_2023,
	title = {Barriers and {Self}-{Efficacy}: {A} {Large}-{Scale} {Study} on the {Impact} of {OSS} {Courses} on {Student} {Perceptions}},
	shorttitle = {Barriers and {Self}-{Efficacy}},
	url = {http://arxiv.org/abs/2304.14628},
	doi = {10.48550/arXiv.2304.14628},
	urldate = {2025-09-15},
	publisher = {arXiv},
	author = {Salerno, Larissa and Tonhão, Simone de França and Steinmacher, Igor and Treude, Christoph},
	month = jul,
	year = {2023},
	note = {arXiv:2304.14628},
}

@misc{carter_skillscope_2025,
  title={Skillscope: A tool to predict fine-grained skills needed to solve issues on github},
  author={Carter, Benjamin C and Contreras, Jonathan Rivas and Villegas, Carlos A Llanes and Acharya, Pawan and Utzerath, Jack and Farner, Adonijah O and Jenkins, Hunter and Johnson, Dylan and Penney, Jacob and Steinmacher, Igor and others},
  booktitle={2025 IEEE/ACM International Workshop on Natural Language-Based Software Engineering (NLBSE)},
  pages={9--12},
  year={2025},
  organization={IEEE}
}

@inproceedings{santos_ossdoorway_2025,
  title={Ossdoorway: A Gamified Environment to Scaffold Student Contributions to Open Source Software},
  author={Santos, Italo and Felizardo, Katia Romero and Sarma, Anita and Steinmacher, Igor and Gerosa, Marco A},
  booktitle={2025 IEEE/ACM 37th International Conference on Software Engineering Education and Training (CSEE\&T)},
  pages={325--335},
  year={2025},
  organization={IEEE Computer Society}
}

@inproceedings{santos_exploring_2025,
  title={Exploring the Untapped: Student Perceptions and Participation in OSS},
  author={Santos, Italo and Felizardo, Katia Romero and Trinkenreich, Bianca and M. German, Daniel and Steinmacher, Igor and A. Gerosa, Marco},
  booktitle={Proceedings of the 33rd ACM International Conference on the Foundations of Software Engineering},
  pages={859--870},
  year={2025}
}

@inproceedings{pinto2019training,
  title={Training software engineers using open-source software: the students' perspective},
  author={Pinto, Gustavo and Ferreira, Clarice and Souza, Cleice and Steinmacher, Igor and Meirelles, Paulo},
  booktitle={2019 IEEE/ACM 41st International Conference on Software Engineering: Software Engineering Education and Training (ICSE-SEET)},
  pages={147--157},
  year={2019},
  organization={IEEE}
}

@inproceedings{holmes2018dimensions,
  title={Dimensions of experientialism for software engineering education},
  author={Holmes, Reid and Allen, Meghan and Craig, Michelle},
  booktitle={Proceedings of the 40th International Conference on Software Engineering: Software Engineering Education and Training},
  pages={31--39},
  year={2018}
}

@INPROCEEDINGS{santos2023designing,
  author={Santos, Italo and Pimentel, João Felipe and Wiese, Igor and Steinmacher, Igor and Sarma, Anita and Gerosa, Marco A.},
  booktitle={IEEE/ACM 45th International Conference on Software Engineering: Software Engineering in Society (ICSE-SEIS)}, 
  title={Designing for cognitive diversity: improving the {GitHub} experience for newcomers}, 
  year={2023},
  pages={1-12}
}

@inproceedings{ellis2008support,
  title={Support for educating software engineers through humanitarian open source projects},
  author={Ellis, H and Morelli, Ralph A and Hislop, GW},
  booktitle={2008 21st IEEE-CS conference on software engineering education and training workshop},
  pages={1--4},
  year={2008},
  organization={IEEE}
}

@inproceedings{santos2022hits,
  title={Hits and Misses: Newcomers' ability to identify skills needed for OSS tasks},
  author={Santos, Italo and Wiese, Igor and Steinmacher, Igor and Sarma, Anita and Gerosa, Marco A},
  booktitle={IEEE SANER},
  pages={174--183},
  year={2022},
  organization={IEEE}
}

@inproceedings{park2009beyond,
  title={Beyond pretty pictures: Examining the benefits of code visualization for open source newcomers},
  author={Park, Yunrim and Jensen, Carlos},
  booktitle={2009 5th IEEE International Workshop on Visualizing Software for Understanding and Analysis},
  pages={3--10},
  year={2009},
  organization={IEEE}
}

@inproceedings{sarma2016training,
  title={Training the future workforce through task Curation in an OSS ecosystem},
  author={Sarma, Anita and Gerosa, Marco Aur{\'e}lio and Steinmacher, Igor and Leano, Rafael},
  booktitle={Proceedings of the 2016 24th ACM SIGSOFT International Symposium on Foundations of Software Engineering},
  pages={932--935},
  year={2016}
}

@article{trinkenreich2021women,
  title={Women's participation in open source software: a survey of the literature},
  author={Trinkenreich, Bianca and Wiese, Igor and Sarma, Anita and Gerosa, Marco and Steinmacher, Igor},
  journal={ACM TOSEM},
  year={2022}
}

@article{steinmacher2015systematic,
  title={A systematic literature review on the barriers faced by newcomers to open source software projects},
  author={Steinmacher, Igor and Silva, Marco Aurelio Graciotto and Gerosa, Marco Aurelio and Redmiles, David F},
  journal={Information and Software Technology (IST)},
  volume={59},
  pages={67--85},
  year={2015},
  publisher={Elsevier}
}

@inproceedings{PS02canfora2012going,
  title={Who is going to mentor newcomers in open source projects?},
  author={Canfora, Gerardo and Di Penta, Massimiliano and Oliveto, Rocco and Panichella, Sebastiano},
  booktitle={ACM SIGSOFT 20th International Symposium on the Foundations of Software Engineering (FSE)},
  pages={1--11},
  year={2012}
}

@article{padala2020gender,
  title={How gender-biased tools shape newcomer experiences in {OSS} projects},
  author={Padala, Susmita Hema and Mendez, Christopher John and Dias, Luiz Felipe and Steinmacher, Igor and Hanson, Zoe Steine and Hilderbrand, Claudia and Horvath, Amber and Hill, Charles and Simpson, Logan Dale and Burnett, Margaret and others},
  journal={IEEE Transactions on Software Engineering (TSE)},
  year={2020},
  publisher={IEEE}
}

@inproceedings{fronchetti2019attracts,
  title={What attracts newcomers to onboard on {OSS} projects? tl; dr: Popularity},
  author={Fronchetti, Felipe and Wiese, Igor and Pinto, Gustavo and Steinmacher, Igor},
  booktitle={15th IFIP Advances in Information and Communication Technology (OSS)},
  pages={91--103},
  year={2019},
  organization={Springer}
}

@misc{von2003special,
  title={Special issue on open source software development},
  author={Von Krogh, Georg and Von Hippel, Eric},
  journal={Research Policy},
  volume={32},
  number={7},
  pages={1149--1157},
  year={2003},
  publisher={Elsevier}
}

\end{document}